\documentclass[a4paper,11pt,dvips]{article}
\usepackage{graphicx}
\usepackage{amsmath}
\usepackage{amssymb}
\usepackage{latexsym}
\usepackage[colorlinks]{hyperref}
\usepackage{color}
\usepackage{float}
\usepackage{cite}
\usepackage{makeidx}
\usepackage{colortbl}

\newcommand{\bra}{\begin{array}}
\newcommand{\era}{\end{array}}
\newcommand{\beq}{\begin{equation}}
\newcommand{\eeq}{\end{equation}}
\newcommand{\bqr}{\begin{eqnarray}}
\newcommand{\eqr}{\end{eqnarray}}

\def\BC{\bb C}
\def\_\BC{\bbi C}

\def \lp {\left(}
\def \gp {\right)}

\def\no2 {{\textstyle{n\over 2}}}

\newcommand{\lb}{\label}

\begin{document}
\begin{titlepage}
\setcounter{page}{1}
\renewcommand{\thefootnote}{\fnsymbol{footnote}}


\vspace{5mm}
\begin{center}

{\Large \bf {Controllable Goos-H\"anchen Shift in Graphene\\
Triangular Double Barrier  }}

\vspace{5mm}
{\bf Miloud Mekkaoui}$^{a}$,
 {\bf Ahmed Jellal\footnote{\sf ajellal@ictp.it --
a.jellal@ucd.ac.ma}}$^{a,b}$ and {\bf Hocine Bahlouli}$^{b,c}$

\vspace{5mm}

{$^{a}$\em Theoretical Physics Group,  
Faculty of Sciences, Choua\"ib Doukkali University},\\
{\em PO Box 20, 24000 El Jadida, Morocco}

{$^b$\em Saudi Center for Theoretical Physics, Dhahran, Saudi Arabia}

{$^c$\em Physics Department,  King Fahd University
of Petroleum $\&$ Minerals,\\
Dhahran 31261, Saudi Arabia}


\vspace{3cm}

\begin{abstract}

We study the Goos-H\"anchen shifts for Dirac fermions in graphene
scattered by a triangular double barrier potential.
The massless Dirac-like equation was used to describe 
the scattered fermions by such potential configuration. Our results show 
that the GHL shifts is affected by the geometrical structure of the double barrier. 
In particular the GHL shifts change sign at the transmission zero
energies and exhibit enhanced peaks at each bound state associated with the double
barrier when the incident angle is less than the critical angle associated with the
total reflection.

\end{abstract}
\end{center}

\vspace{3cm}


\end{titlepage}


\section{Introduction}


During the few past years there is a progress in studying electron 
transport properties in the graphene systems~\cite{Novoselov}. With this respect, we cite the quantum version of
the Goos-H\"anchen  effect originating from the reflection of particles from
interfaces.  The Goos-H\"anchen shift was discovered by Hermann Fritz Gustav
Goos and Hilda H\"anchen~\cite{FGoos, FFGoos} and theoretically
explained by Artman~\cite{Artmann} in the late of 1940s. Many
works in various graphene-based nanostructures, including
single~\cite{Chen15}, double barrier~\cite{Song16} and
superlattices~\cite{Chen18}, showed that the Goos-H\"anchen like
(GHL) shifts can be enhanced by the transmission resonances and
controlled by varying the electrostatic potential and induced
gap~\cite{Chen15}. Similar to those in semiconductors, the GHL
shifts in graphene can also be modulated by the electric and magnetic
barriers~\cite{Sharma19} as well as atomic optics~\cite{Huang13}. It has
been reported that the GHL shifts play an important role in the
group velocity of quasiparticles along interfaces of graphene p-n
junctions~\cite{Beenakker,Zhao11}. 

Very recently, we have studied the Dirac
fermions in graphene scattered by a triangular double barrier
in terms of the transmission probability~\cite{MMekkaoui}.
The system was made of two triangular potential barrier regions separated by a well region 
characterized by an energy gap $G_p$. Solving the Dirac-like equation and matching the solutions 
at the boundaries, the transmission and reflection coefficients were expressed in terms of transfer matrix. 
In particular, it is showed that the transmission exhibits 
oscillation resonances that are manifestation of the Klein tunneling effect.


Actually we are wondering to extend our work~\cite{MMekkaoui} to deal with
other issues related to graphene systems. Indeed, 
we investigate 
the GHL shifts for a system made of graphene with 
gap an in presence 
of the triangular double barrier potential. By
splitting our system into three regions, we determine the
solutions of the energy spectrum in terms of different physical
quantities. After matching the wave functions at both interfaces, 
we calculate the transmission coefficient as well as the GHL shifts. 
To give a better understanding of our results, we plot the GHL shifts versus various 
physical parameters characterizing our system. 

The paper is organized as follows. In section 2, we formulate our
model by setting the Hamiltonian system describing particles
scattered by a triangular double barrier whose intermediate zone is
subject to a mass term. In section 3,  we obtain the spinor
solution corresponding to each regions composing our system.  We
use the transfer matrix to describe the boundary conditions and
split the energy regions into three domains in order to calculate the the
phase shift and GHL shifts. In section 4, we numerically present
our results for the GH shifts and the transmission probability of
an electron beam transmitted through a graphene triangular double barrier.
We then conclude our work in the final section.

\section{System model}

We consider a system of massless Dirac fermions through a strip of
graphene with the Fermi energy $E$ and the incidence angle
$\phi_1$ with respective to the incident $x$-direction of two-dimensional
graphene sheet subject to a triangular double barrier potential. Specifically this system
is a flat sheet of graphene subject to a square potential barrier
along the $x$-direction while particles are free in the
$y$-direction. For ease of mathematical formulation let us first describe the
geometry of our system as being made of five regions denoted by
${\sf j}$ = $1, \cdots, 5$. Each region is characterized by its
potential and interaction with external sources. The barrier
regions are formally described by a Dirac-like Hamiltonian
\begin{equation}\lb{Ham1}
H=v_{F}
{\boldsymbol{\sigma}}\cdot\textbf{p}+
V(x){\mathbb
I}_{2}+\Delta\sigma_{z}
\end{equation}
where
${v_{F}\approx 10^6 m/s}$  is the Fermi velocity,
${{\boldsymbol{\sigma}}=(\sigma_{x},\sigma_{y})}$ are the Pauli
matrices, $\textbf{p}=-i\hbar(\partial_{x},
\partial_{y})$, ${\mathbb I}_{2}$ the $2 \times 2$ unit matrix,
the electrostatic potential $V(x)=V_{\sf j}$ in each scattering region. 
The parameter $\Delta = m v_{F}^2$ is the energy gap originating either from 
sublattice symmetry breaking or from the spin-orbit interaction. It 
is defined by
\begin{equation}
\Delta=t'\Theta\left(d_{1}^{2}-x^{2}\right)
\end{equation}
where  $\Theta$ is the Heaviside step function, $d_{1}$ and $t'$ are positive numbers 
defining the width and strength of the energy gap region.
In order to study the scattering of Dirac fermions in graphene by the above 
double barrier structure we first choose the following explicit potential configuration
\begin{equation}\lb{poten}
V(x)=V_{\sf j}=
\left\{%
\begin{array}{ll}
    (\gamma x+d_2)F, & \hbox{$d_{1}\leq |x|\leq d_{2}$} \\
    V_{2}, & \hbox{$ |x|\leq d_{1}$} \\
    0, & \hbox{otherwise} \\
\end{array}%
\right.
\end{equation}
with $d_{1}$ is a positive number such that $d_{2}-d_{1}$ represents the width 
of the triangular potential barrier region and 
$\gamma=\pm1$,  $\gamma=1$ for $x\in [-d_2, -d_1]$,
$\gamma=-1$ for $x\in [d_1, d_2]$ and $F=\frac{v_1}{d_2-d_1}$.

\begin{figure}[!ht]
  \centering
  \includegraphics[width=8cm, height=4cm ]{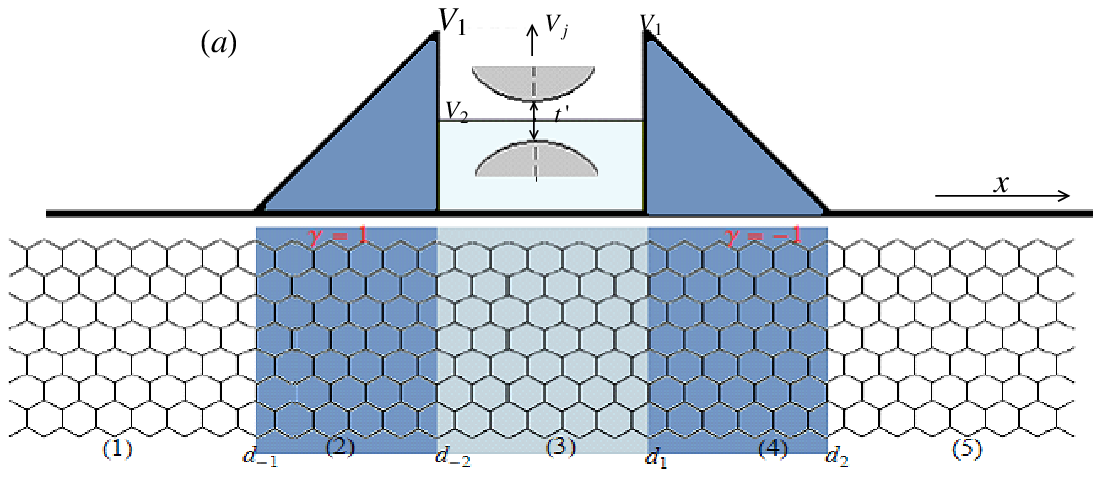}\ \ \ \ \ \ \
  \includegraphics[width=7cm, height=4cm ]{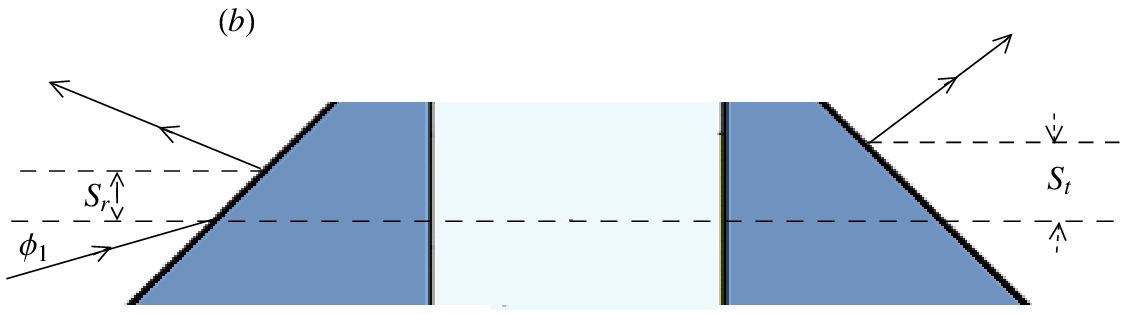}
  \caption{\sf Schematic diagram for Dirac fermions in
inhomogeneous magnetic field through a graphene double barrier,
with the two barriers triangular of width $d_{2}-d_{1}$ height
$V_{1}$, and
  distance $2d_{1}$ height potential $V_{2}$ between them. (a): the dashed lines show smooth electric potentials
with distributions of error functions. (b): describes the incident,
reflected, and transmitted electron beams with a lateral shift
$S_{r}$ and $S_{t}$.}\label{db.1}
\end{figure}
We define each potential region as follows: ${\sf j = 1}$ for $x \leq -d_2
$, ${\sf j = 2}$ for $ -d_2 \leq x \leq -d_1 $, ${\sf j = 3}$ for $ -d_1 \leq x
\leq d_1 $, ${\sf j = 4}$ for $ d_1 \leq x \leq d_2 $ and ${\sf j = 5}$ for $ x
\geq d_2 $. The corresponding constant potentials are given in \eqref{poten}
and are denoted by $V_{\sf j}$ in the ${\sf j}$-th region,  the
five regions indicated schematically in Figure \ref{db.1}, which
shows the space configuration of the potential profile. We
therefore need to study just one \textbf{K} point. The
time-independent Dirac equation for the spinor
$\Phi(x,y)=\left(\varphi^{+},\varphi^{-}\right)^{T}$, where $T$ stands for 
the transpose and $E=v_{F}\epsilon$ is the energy of the system, which can be defined by 
\begin{equation} \lb{eqh1}
\left[{\boldsymbol{\sigma}}\cdot\textbf{p}+ v_{\sf j}{\mathbb
I}_{2}+\mu\Theta\left(d_{1}^{2}-x^{2}\right)\sigma_{z}\right]\Phi(x,y)=\epsilon
\Phi(x,y)
\end{equation}
in the unit system $\hbar=1$, with
$V_{\sf j}=v_{F}v_{\sf j}$, $t'=v_{F}\mu$, $F=v_{F}\varrho$
and $V_{1}=v_{F}v_{1}$.
Our system is supposed to have finite width $W$ with infinite mass
boundary conditions on the wavefunction at the boundaries $y = 0$
and $y = W$ along the $y$-direction \cite{Tworzydlo, Berry}. These
boundary conditions result in a quantization of the transverse
momentum along the $y$-direction 
\begin{equation}
k_{y}=\frac{\pi}{W}\left(n+\frac{1}{2}\right),\qquad n=0,1,2 \cdots.
\end{equation}

One can therefore assume a spinor solution of the 
form
$\Phi_{\sf j}=\left(\varphi_{\sf j}^{+}(x),\varphi_{\sf
j}^{-}(x)\right)^{T}e^{ik_{y}y}$ and the subscripts ${\sf
j}= 1, 2, 3, 4, 5$, indicates the space region while the
superscripts indicate the two spinor components. Solving the
eigenvalue equation to obtain the upper and lower components of
the eignespinor in the incident and reflection region ($x < - d_{2}$)
\begin{equation}\label{eq3}
    \Phi_{\sf 1}=  \left(
            \begin{array}{c}
              {1} \\
              {z_{1}} \\
            \end{array}
          \right) e^{i(k_{1}x+k_{y}y)} + r_{s}\left(
            \begin{array}{c}
              {1} \\
              {-z_{1}^{-1}} \\
            \end{array}
          \right) e^{i(-k_{1}x+k_{y}y)}
\end{equation}
\beq
z_{1} =s_{1}\frac{k_{1}+ik_{y}}{\sqrt{k_{1}^{2}+k_{y}^{2}}}
\eeq
where the
sign function is defined by $s_{\sf j}={\mbox{sign}}{\left(E\right)}$.
The corresponding dispersion relation is given by \beq
\epsilon=s_1\sqrt{k_1^2 +k_y^2}.
\eeq
In region {\sf 2} and {\sf 4} ($d_{1}<|x|<d_{2}$),  the general
solution can be expressed in terms of the parabolic cylinder
function \cite{Abramowitz, Gonzalez, HBahlouli} as
\begin{equation}\lb{hiii1}
 \chi_{\gamma}^{+}=c_{n1}
 D_{\nu_n-1}\left(Q_{\gamma}\right)+c_{n2}
 D_{-\nu_n}\left(-Q^{*}_{\gamma}\right)
\end{equation}
where $\nu_n=\frac{ik_{y}^{2}}{2\varrho}$, $\epsilon_{0}=\epsilon-v_{1}$, $
Q_{\gamma}(x)=\sqrt{\frac{2}{\varrho}}e^{i\pi/4}\left(\gamma
\varrho x+\epsilon_{0}\right) $, $c_{n1}$ and $c_{n2}$ are
constants. The second component  reads as
\begin{eqnarray}\lb{hiii2}
\chi_{\gamma}^{-}=-\frac{c_{n2}}{k_{y}}\left[
2(\epsilon_{0}+\gamma \varrho x)
 D_{-\nu_n}\left(-Q^{*}_{\gamma}\right)
+
 \sqrt{2\varrho}e^{i\pi/4}D_{-\nu_n+1}\left(-Q^{*}_{\gamma}\right)\right]
 -\frac{c_{n1}}{k_{y}}\sqrt{2\varrho}e^{-i\pi/4}
 D_{\nu_n-1}\left(Q_{\gamma}\right)
\end{eqnarray}
The components of the spinor solution of the Dirac equation
\eqref{eqh1} in region {\sf 2} and {\sf 4} can be obtained from
\eqref{hiii1} and \eqref{hiii2} with
$\varphi_{\gamma}^{+}(x)=\chi_{\gamma}^{+}+i\chi_{\gamma}^{-}$ and
$\varphi_{\gamma}^{-}(x)=\chi_{\gamma}^{+}-i\chi_{\gamma}^{-}$. In
regions {\sf j}={\sf 2,4} 
we have the eigenspinors
\begin{eqnarray}
 \Phi_{\sf j } &=& a_{\sf j-1}\left(%
\begin{array}{c}
 \eta^{+}_{\gamma}(x) \\
  \eta^{-}_{\gamma}(x) \\
\end{array}%
\right)e^{ik_{y}y}+a_{\sf j}\left(%
\begin{array}{c}
 \xi^{+}_{\gamma}(x) \\
 \xi^{-}_{\gamma}(x)\\
\end{array}%
\right)e^{ik_{y}y}
\end{eqnarray}
The function $ \eta^{\pm}_{\gamma}(x)$ and $\xi^{\pm}_{\gamma}(x)$
are given by
\begin{eqnarray}
\eta^{\pm}_{\gamma}(x) &=&
 D_{\nu_{n}-1}\left(Q_{\gamma}\right)\mp
 \frac{1}{k_{y}}\sqrt{2\varrho}e^{i\pi/4}D_{\nu_{n}}\left(Q_{\gamma}\right)\\
\xi^{\pm}_{\gamma}(x)&=&
 \pm\frac{1}{k_{y}}\sqrt{2\varrho}e^{-i\pi/4}D_{-\nu_{n}+1}\left(-Q_{\gamma}^{*}\right)\nonumber\\
 &&
  \pm
 \frac{1}{k_{y}}\left(-2i\epsilon_{0}\pm
 k_{y}-\gamma2i \varrho x\right)D_{-\nu_{n}}\left(-Q_{\gamma}^{*}\right).
\end{eqnarray}
In region {\sf 2}:
\begin{eqnarray}
 \Phi_{\sf 2} &=& a_1\left(%
\begin{array}{c}
 \eta^{+}_{1}(x) \\
  \eta^{-}_{1}(x) \\
\end{array}%
\right)e^{ik_{y}y}+a_{2}\left(%
\begin{array}{c}
 \xi^{+}_{1}(x) \\
 \xi^{-}_{1}(x)\\
\end{array}%
\right)e^{ik_{y}y}
\end{eqnarray}
In region {\sf 4}:
\begin{eqnarray}
 \Phi_{\sf 4} &=& a_3\left(%
\begin{array}{c}
 \eta^{+}_{-1}(x) \\
  \eta^{-}_{-1}(x) \\
\end{array}%
\right)e^{ik_{y}y}+a_{4}\left(%
\begin{array}{c}
 \xi^{+}_{-1}(x) \\
 \xi^{-}_{-1}(x)\\
\end{array}%
\right)e^{ik_{y}y}
\end{eqnarray}
where $\gamma=\pm 1$. Solving the eigenvalue equation for
the Hamiltonian \eqref{eqh1} describing region 3, we find the
following eigenspinor
\begin{equation} \label{eq 7}
 \Phi_{\sf 3}= b_1 \left(
            \begin{array}{c}
              {\alpha} \\
              {\beta z_{3}} \\
            \end{array}
          \right) e^{i(k_{3}x+k_{y}y)} +b_2 \left(
            \begin{array}{c}
              {\alpha} \\
              {-\beta z_{3}^{-1}} \\
            \end{array}
          \right) e^{i(-k_{3}x+k_{y}y)}
\end{equation}
with the parameters $\alpha$ and $\beta$ are defined by
\begin{equation} \label{eq 18i}
       {\alpha=\lp{1+\frac{\mu}{ \epsilon-v_{2}}}\gp}^{1/2}, \qquad
       {\beta=\lp{1-\frac{\mu}{ \epsilon-v_{2}}}\gp}^{1/2}
 \end{equation}
and the complex number  
\beq z_{3}
=s_{3}\frac{k_{3}+ik_{y}}{\sqrt{k_{3}^{2}+k_{y}^{2}}} 
\eeq 
with the sign function $s_{3}=\mbox{sign}(\epsilon-v_{2})$.
The wave vector being
\beq
k_{3}= \sqrt{(\epsilon-v_{2})^{2}-\mu^{2}-{k_{y}}^{2}}
\eeq
Finally the eigenspinor in region {\sf 5} ($x
> d_{2}$) can be expressed as
\begin{equation}\label{eq6}
 \Phi_{\sf 5}= t_{s} \left(
            \begin{array}{c}
              {1} \\
              {z_{1}} \\
            \end{array}
          \right) e^{i(k_{1}x+k_{y}y)}.
\end{equation}

In the next, we will see how to use the above solutions in order to deal with different issues. These
concern the transmission and reflection probabilities, which will allow us 
to determine the phase shift and therefore study the Goos-H\"anchen like shifts.

\section{ Phase shift and GHL shifts}

The transmission and reflection coefficients $(r_{s},t_{s})$
can be determined using the boundary conditions and continuity of
the eigenspinors at each interface. These will help to 
build a bridge between quantum optics and Dirac
fermions in graphene through the GHL shifts. We prefer to express these relationships in
terms of $2\times 2$
transfer matrices between different regions $M_{j,j+1}$, such as
\beq
\left(%
\begin{array}{c}
  a_{\sf j} \\
  b_{\sf j} \\
\end{array}%
\right)=M_{\sf j, j+1}\left(%
\begin{array}{c}
  a_{\sf j+1} \\
  b_{\sf j+1} \\
\end{array}%
\right).
\eeq
From this, we finally end up with the full transfer matrix over the
whole double barrier which can be written, in an obvious notation,
as
\begin{equation}\label{systm1}
\left(%
\begin{array}{c}
  1 \\
  r_{s} \\
\end{array}%
\right)=\prod_{\sf j=1}^{4}M_{\sf j j+1}\left(%
\begin{array}{c}
  t_{s} \\
  0 \\
\end{array}%
\right)=M\left(%
\begin{array}{c}
  t_{s} \\
  0 \\
\end{array}%
\right)
\end{equation}
where the total transfer matrix $M=M_{12}\cdot M_{2
3}\cdot M_{34}\cdot M_{45}$ and $M_{\sf j j+1}$ are transfer matrices that couple the
wave function in the $\sf j$-th region to the wave function in the
$(\sf{j+ 1})$-th region. These are given explicitly by
\begin{eqnarray}
&& M=\left(%
\begin{array}{cc}
  m_{11} & m_{12} \\
  m_{21} & m_{22} \\
\end{array}%
\right)
\\
&& M_{12}=\left(%
\begin{array}{cc}
   e^{-\textbf{\emph{i}}k_{1} d_{2}} &e^{\textbf{\emph{i}}k_{1} d_{2}} \\
  z_{1}e^{-\textbf{\emph{i}}k_{1} d_{2}} & -z^{\ast}_{1} e^{\textbf{\emph{i}}k_{1} d_{2}} \\
\end{array}%
\right)^{-1}\left(%
\begin{array}{cc}
\eta_{1}^{+}(-d_2) &  \xi_{1}^{+}(-d_2)\\
 \eta_{1}^{-}(-d_2) & \xi_{1}^{-} (-d_2)\\
\end{array}%
\right)
\\
&& M_{23}=\left(%
\begin{array}{cc}
 \eta_{1}^{+}(-d_1) &  \xi_{1}^{+}(-d_1)\\
 \eta_{1}^{-}(-d_1) & \xi_{1}^{-} (-d_1)\\
\end{array}%
\right)^{-1}\left(%
\begin{array}{cc}
\alpha e^{-\textbf{\emph{i}}k_{3} d_{1}} &\alpha e^{\textbf{\emph{i}}k_{3} d_{1}} \\
  \beta z_{3}e^{-\textbf{\emph{i}}k_{3} d_{1}} & -\beta z^{\ast}_{3} e^{\textbf{\emph{i}}k_{3} d_{1}} \\
\end{array}%
\right)
\\
&& M_{34}=\left(%
\begin{array}{cc}
 \alpha e^{\textbf{\emph{i}}k_{3} d_{1}} &\alpha e^{-\textbf{\emph{i}}k_{3} d_{1}} \\
  \beta z_{3}e^{\textbf{\emph{i}}k_{3} d_{1}} & -\beta z^{\ast}_{3} e^{-\textbf{\emph{i}}k_{3} d_{1}} \\
\end{array}%
\right)^{-1}\left(%
\begin{array}{cc}
 \eta_{-1}^{+}(d_1) &  \xi_{-1}^{+}(d_1)\\
 \eta_{-1}^{-}(d_1) & \xi_{-1}^{-} (d_1)\\
\end{array}%
\right)
\\
&& M_{45}=\left(%
\begin{array}{cc}
 \eta_{-1}^{+}(d_2) &  \xi_{-1}^{+}(d_2)\\
 \eta_{-1}^{-}(d_2) & \xi_{-1}^{-} (d_2)\\
\end{array}%
\right)^{-1}\left(%
\begin{array}{cc}
  e^{\textbf{\emph{i}}k_{1} d_{2}} & e^{-\textbf{\emph{i}}k_{1} d_{2}} \\
  z_{1} e^{\textbf{\emph{i}}k_{1} d_{2}}  & -z_{1}^{\ast} e^{-\textbf{\emph{i}}k_{1} d_{2}}  \\
\end{array}%
\right)
\end{eqnarray}
with the following relationships between the parabolic cylindrical functions 
\begin{eqnarray}
&&\eta_{-1}^{\pm}(d_1)=\eta_{1}^{\pm}(-d_1),\qquad
\eta_{-1}^{\pm}(d_2)=\eta_{1}^{\pm}(-d_2)
 \\
 &&
 \xi_{-1}^{\pm}(d_1)=\xi_{1}^{\pm}(-d_1),\qquad
 \xi_{-1}^{\pm}(d_2)=\xi_{1}^{\pm}(-d_2).
\end{eqnarray}
The above analysis allows
to extract the transmission and reflection amplitudes as
\begin{equation}\label{eq 63}
 t_{s}=\frac{1}{m_{11}}, \qquad  r_{s}=\frac{m_{21}}{m_{11}}.
\end{equation}

At this stage, we should point out  that we were unfortunately
forced to adopt a somehow cumbersome notation for our wavefunction
parameters in different potential regions due to the relatively
large number of necessary subscripts and superscripts. Before
matching the eigenspinors at the boundaries, let us define the
following shorthand notation
\begin{eqnarray}
&& \eta_{1}^{\pm}(-d_1)=\eta_{11}^{\pm},\qquad
 \eta_{1}^{\pm}(-d_2)=\eta_{12}^{\pm}\\
 &&
 \xi_{1}^{\pm}(-d_1)=\xi_{11}^{\pm},\qquad
 \xi_{1}^{\pm}(-d_2)=\xi_{12}^{\pm}
\end{eqnarray}
Now we are able to explicitly
determine the transmission amplitude $t_{s}$. Indeed,  after
some lengthy algebra, one can solve the linear system given in
\eqref{systm1} to obtain the transmission and reflection
amplitudes in closed form. We obtain
\begin{equation}
t_{s}=\frac{\alpha\beta e^{2i(k_{1}d_{2}+k_{3}d_{1})}
\left(1+z_{1}^{2}\right)\left(1+z_{3}^{2}\right)}{z_{3}\left(e^{4ik_{3}d_{1}}-1\right)\left(
\alpha^{2}G_{2}+\beta^{2}G_{1}\right)+\alpha\beta
G_{3}}\left(\xi_{11}^{+}\eta_{11}^{-}-\xi_{11}^{-}\eta_{11}^{+}\right)
\left(\xi_{12}^{-}\eta_{12}^{+}-\xi_{12}^{+}\eta_{12}^{-}\right)
 \end{equation}
where we have defined the following quantities
\begin{eqnarray}
 &&G_{1}=\left(\xi_{12}^{-}\eta_{11}^{+}-\xi_{11}^{+}\eta_{12}^{-}-
 \xi_{12}^{+}\eta_{11}^{+}z_{1}+\xi_{11}^{+}\eta_{12}^{+}z_{1}\right)\left( \xi_{11}^{+}\eta_{12}^{+}+
 \xi_{11}^{+}\eta_{12}^{-}z_{1}-\eta_{11}^{+}(\xi_{12}^{+}+\xi_{12}^{-}z_{1}\right)\\
 &&
 G_{2}=\left(\xi_{11}^{-}\eta_{12}^{+}-\xi_{11}^{-}\eta_{12}^{-}z_{1}-\eta_{11}^{-}(
 \xi_{12}^{+}+\xi_{12}^{+}z_{1}\right)
 \left( -\xi_{12}^{-}\eta_{11}^{-}+
 \xi_{12}^{+}\eta_{11}^{-}z_{1}-\xi_{11}^{-}(\eta_{12}^{-}+\eta_{12}^{+}z_{1}\right)\\
 &&
 G_{3}=\Gamma_{0}\left(1+z_{1}^{2}z_{3}^{2}\right)+\Gamma_{1}z_{1}\left(1-z_{3}\right)+\Gamma_{2}\left(z_{1}^{2}+z_{3}^{2}\right)
  +e^{4id_{1}k_{3}}\left(\Gamma_{3}+\Gamma_{4}\right)
\end{eqnarray}
as well as
 \begin{eqnarray}
 \Gamma_{0}&=&-\xi_{12}^{+}\xi_{12}^{-}\eta_{11}^{+}\eta_{11}^{-}
 +\xi_{11}^{+}\xi_{12}^{-}\eta_{11}^{-}\eta_{12}^{+}+
 \xi_{11}^{-}\xi_{12}^{+}\eta_{11}^{+}\eta_{12}^{-}-
 \xi_{11}^{+}\xi_{11}^{-}\eta_{12}^{+}\eta_{12}^{-}\\
 \Gamma_{1}&=&\left(\xi_{12}^{+}\right)^{2}\eta_{11}^{+}\eta_{11}^{-}
 -\left(\xi_{12}^{-}\right)^{2}\eta_{11}^{+}\eta_{11}^{-}-
 \xi_{11}^{-}\xi_{12}^{+}\eta_{11}^{+}\eta_{12}^{+}-
 \xi_{11}^{+}\xi_{12}^{+}\eta_{11}^{-}\eta_{12}^{+}\\\nonumber
 &&
 +\xi_{11}^{+}\xi_{11}^{-}\left(\eta_{12}^{+}\right)^{2}
 -\xi_{11}^{+}\xi_{11}^{-}\left(\eta_{12}^{-}\right)^{2}+
 \xi_{11}^{-}\xi_{12}^{-}\eta_{11}^{+}\eta_{12}^{-}+
 \xi_{11}^{+}\xi_{12}^{-}\eta_{11}^{-}\eta_{12}^{-}\\
 \Gamma_{2}&=&\xi_{12}^{+}\xi_{12}^{-}\eta_{11}^{+}\eta_{11}^{-}
 -\xi_{11}^{-}\xi_{12}^{-}\eta_{11}^{+}\eta_{12}^{+}-
 \xi_{11}^{+}\xi_{12}^{+}\eta_{11}^{-}\eta_{12}^{-}+
 \xi_{11}^{+}\xi_{11}^{-}\eta_{12}^{+}\eta_{12}^{-}\\
 \Gamma_{3}&=&\left(\xi_{12}^{+}\right)^{2}\eta_{11}^{+}\eta_{11}^{-}\left(z_{3}^{2}-1\right)
 -\xi_{11}^{-}\xi_{12}^{-}\eta_{11}^{+}\left[\eta_{12}^{+}\left(1+z_{1}^{2}z_{3}^{2}\right)-\eta_{12}^{-}z_{1}\left(z_{3}^{2}-1\right)\right]\\\nonumber
 &&
 +\xi_{11}^{-}\xi_{11}^{+}\left[\left(\eta_{12}^{+}\right)^{2}z_{1}
 -\left(\eta_{12}^{-}\right)^{2}z_{1}+\eta_{12}^{+}\eta_{12}^{-}\left(z_{1}^{2}-1\right)\left(z_{3}^{2}-1\right)\right]
 \\
 \Gamma_{4}&=&\xi_{12}^{-}\eta_{11}^{-}\left[-\xi_{12}^{-}\eta_{11}^{+}z_{1}\left(z_{3}^{2}-1\right)+
 \xi_{11}^{+}\left(\eta_{12}^{-}z_{0}\left(z_{3}^{2}-1\right)+\eta_{12}^{+}\left(z_{1}^{2}+z_{3}^{2}\right)\right)\right]\\\nonumber
 &&\xi_{12}^{+}\xi_{12}^{-}\eta_{11}^{+}\eta_{11}^{-}\left(z_{1}^{2}+1\right)\left(z_{1}^{3}-1\right)-\xi_{12}^{+}
 \xi_{11}^{+}\eta_{11}^{-}\left(\eta_{12}^{-}\left(1+z_{1}^{2}z_{3}^{2}\right)+\eta_{12}^{+}z_{1}\left(z_{1}^{3}-1\right)\right)\\\nonumber
 &&
 +\xi_{12}^{+}\xi_{11}^{-}\eta_{11}^{+}\left[\eta_{12}^{-}\left(z_{1}^{2}+z_{3}^{2}\right)+\eta_{12}^{+}z_{1}\left(1-z_{3}^{2}\right)\right]
\end{eqnarray}
The transmission  and reflection amplitudes can be expressed 
as complex numbers
\begin{eqnarray}
 && t_{s} = \rho_{t_{s}} e^{i\varphi_{t_{s}}}
\\
&&
r_{s} = \rho_{r_{s}}e^{i\varphi_{r_{s}}}
\end{eqnarray}
where we have defined the phase shift of the transmission and reflection amplitudes as being 
$\varphi_{t}$ and $\varphi_{r}$, respectively
\begin{equation}
     \varphi_{t_{s}} = \arctan\left(\frac{\Im [t_{s}]}{\Re [t_{s}]}\right),
     \qquad \varphi_{r_{s}} = \arctan\left(\frac{\Im [r_{s}]}{\Re [r_{s}]}\right)
 \end{equation}
 as well as the amplitudes
 \begin{equation}
     \rho_{t_{s}} =\left(\Re [t_{s}]^{2}+\Im [t_{s}]^{2}\right)^{\frac{1}{2}},
     \qquad \rho_{r_{s}} =\left(\Re [r_{s}]^{2}+\Im [r_{s}]^{2}\right)^{\frac{1}{2}}
 \end{equation}
 with as usual the real and imaginary parts are given by
\begin{eqnarray}
&&\Re [t_{s}]=\frac{t_{s}+t_{s}^{\ast}}{2},\qquad
 \Im [t_{s}]=\frac{t_{s}-t_{s}^{\ast}}{2i}
\\
&& \Re [r_{s}]=\frac{r_{s}+r_{s}^{\ast}}{2},\qquad
 \Im [r_{s}]=\frac{r_{s}-r_{s}^{\ast}}{2i}.
\end{eqnarray}
These allow us to express the
 phase shift as
 \begin{equation}
     \varphi_{t_{s}} = \arctan\left(i\frac{t_{s}^{\ast}-t_{s}}{t_{s}+t_{s}^{\ast}}\right),
     \qquad \varphi_{r_{s}} = \arctan\left(i\frac{r_{s}^{\ast}-r_{s}}{r_{s}+r_{s}^{\ast}}\right).
 \end{equation}

We study GHL shifts in graphene by considering an
incident, reflected and transmitted beams with some transverse
wave vector $k_y = k_{y_0}$ and angle of incidence
$\phi_{1}(k_{y_{0}})\in [0, \frac{\pi}{2}]$, denoted by the
subscript $0$. These can be expressed for the incident beam as 
\begin{eqnarray}
   \Psi_{in}(x,y) &=& \int_{-\infty}^{+\infty}dk_y\ f(k_y-k_{y_0})\ e^{i(k_{x1}(k_y)x+k_yy)}\left(
            \begin{array}{c}
              {1} \\
              {e^{i\phi_{1}(k_{y})}}
            \end{array}
          \right)\label{eq 79}
\end{eqnarray}
as well as the reflected beam 
\begin{eqnarray}
\Psi_{re}(x,y) &=& \int_{-\infty}^{+\infty}dk_y\ r_{s}(k_y)\
f(k_y-k_{y_0})\ e^{i(-k_{x1}(k_y)x+k_yy)}\left(
            \begin{array}{c}
              {1} \\
              {-e^{-i\phi_{1}(k_{y})}} \\
            \end{array}
          \right)\label{refl}
\end{eqnarray}
and
the reflection amplitude is
\beq
r_{s}(k_y)=|r_{s}|e^{i\varphi_{r_{s}}}
\eeq
this
fact is represented by writing the $x$-component of wave vector
$k_{x1}$ as well as $\phi_{1}$ both as function of $k_{y}$, where
each spinor plane wave is a solution of \eqref{eqh1}. The function
$f(k_y-k_{y_0})$ is the angular spectral distribution, which can
be assumed of Gaussian shape
\begin{equation}
f(k_y-k_{y_0})=w_ye^{-w_{y}^2(k_y-k_{y_0})^2}
\end{equation}with $w_y$ is the half beam width at waist
\cite{Beenakker}. We can approximate the $k_{y}$-dependent terms
by a Taylor expansion around $k_{y}$ and retaining only the first
order term to get for the phase
\begin{equation}
\phi_{1}(k_{y})\approx
\phi_{1}(k_{y_{0}})+\frac{\partial\phi_{1}}{\partial
k_{y}}\Big|_{k_{y_{0}}}(k_{y}-k_{y_{0}})
\end{equation}
and also for the wave vector
\begin{equation}
k_{x1}(k_{y})\approx k_{x1}(k_{y_{0}})+\frac{\partial
k_{x1}}{\partial k_{y}}\Big|_{k_{y_{0}}}(k_{y}-k_{y_{0}}).
\end{equation}
As far as
the transmission waves is concerned, we write the beam as 
\begin{eqnarray}
\Psi_{tr}(x,y) &=& \int_{-\infty}^{+\infty}dk_y\ t_{s}(k_y)\
f(k_y-k_{y_0})\ e^{i(k_{x1}(k_y)x+k_yy)}\left(
            \begin{array}{c}
              {1} \\
              {e^{i\phi_{1}(k_{y})}} \\
            \end{array}
          \right) \label{trans}
\end{eqnarray}
 and the transmission amplitude is
 \beq
t_{s}(k_y)=|t_{s}|e^{i\varphi_{t_{s}}}
\eeq
which will be
calculated through the use of the boundary conditions. The stationary-phase
approximation indicates that  the GHL shifts are equal to the
negative gradient of transmission phase with respect to $k_y$.
 To  calculate  the  GHL  shifts  of  the
transmitted beam through our system, according to the stationary
phase method \cite{Bohm}, we adopt the definition \cite{Chen15,
AAJellal, AAAJellal}
 \begin{equation}
        S_{t}=- \frac{\partial \varphi_{t_{s}}}{\partial
        k_{y}}\Big|_{k_{y0}}, \qquad S_{r}=- \frac{\partial \varphi_{r_{s}}}{\partial
        k_{y}}\Big|_{k_{y0}}.
 \end{equation}
 These can be used to write the reflection and transmission probabilities as
\begin{equation}
  T_{s}= |t_{s}|^{2},  \qquad
  R_{s}=|r_{s}|^{2}.
\end{equation}
Obviously, we can
check that the probability conservation condition $T_{s}+R_{s}=1$ is well
satisfied. Having obtained the closed form expressions of the GHL shifts and transmission in different energy domains, we
proceed now to compute these quantities numerically. This will help us understand the effect of various
potential parameters on the GHL shifts in our double linear barrier potential.

\section{Discussion of numerical results} 

   The physics of particle scattering through a linear double 
barrier depends on the energy of the incoming particle.
We numerically evaluate the GHL shifts in transmission $S_{t}$ and in reflection $S_{r}$ as a
function of structural parameters of the graphene double linear barrier, including the energy
$\epsilon$, the $y$-component of the wave vector $k_{y}$, the energy gap $\mu$ and the potentials 
$v_{1}$ and $v_{2}$. To understand their behaviors, let us consider Figure \ref{fig3}(a) where
we study the GHL shifts in transmission as well as the GHL shifts in reflection
versus the energy $\epsilon$ for specific values of the parameters
$d_{1}=1$, $d_{2}=2.5$, $v_{1} =60$, $v_{2} =30$. It was found that
the GHL shifts can be negative as well as positive and become zero
at transmission resonances in Figure \ref{fig3}(b). However, in 
intervals $\epsilon\leq v_{2}-k_{y}$ and
$v_{2}+k_{y}\leq\epsilon\leq v_{1}$, there are oscillation
resonances due to the Klein regime, that is, a situation in which
only oscillatory solutions exist throughout and where the so
called Klein paradox reigns. Finally in the interval where 
$\epsilon> v_{1}$ the usual high energy barrier
oscillations and asymptotically the transmission reaches unity at high energy.

\begin{figure}[!ht]
\centering
\includegraphics[width=8cm, height=5cm]{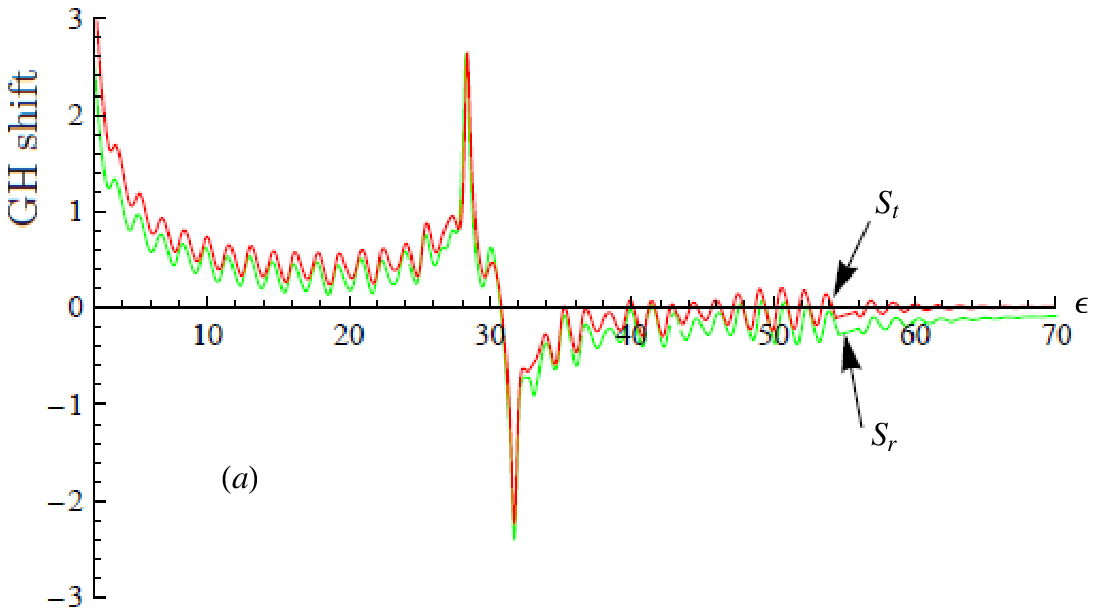}\ \ \ \
\includegraphics[width=8cm, height=5cm]{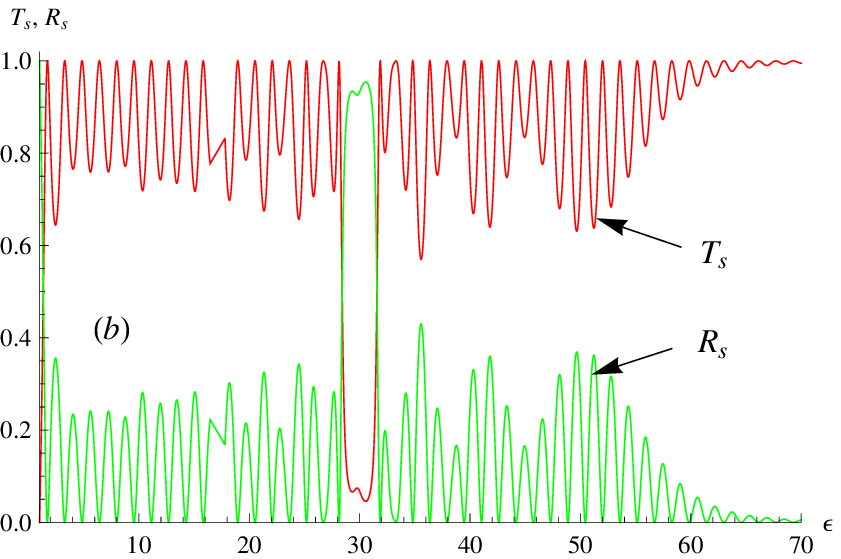}\\
 \caption{\sf{(Color online) (a): GHL shifts in transmission $S_{t}$ and in reflection $S_{r}$ as a function of energy
 $\epsilon$. (b): Transmission and reflection probabilities $(T_{s}, R_{s})$ as a function of energy
 $\epsilon$ with $d_{1}=1$, $d_{2}=2.5$, $\mu=0$,
 $k_{y}=1$,
 $v_{1}=60$ and $v_{2}=30$.}}\lb{fig3}
\end{figure}

\begin{figure}[!ht]
\centering
\includegraphics[width=8cm, height=5cm]{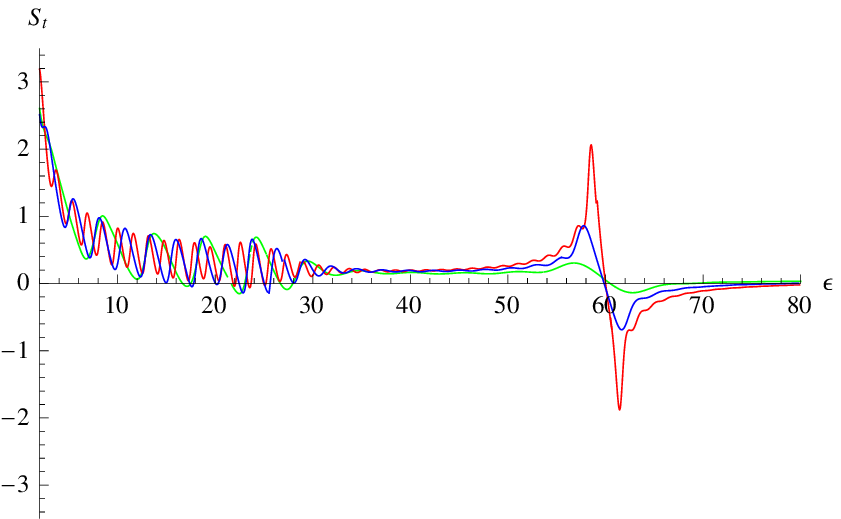}\ \ \ \
\includegraphics[width=8cm, height=5cm]{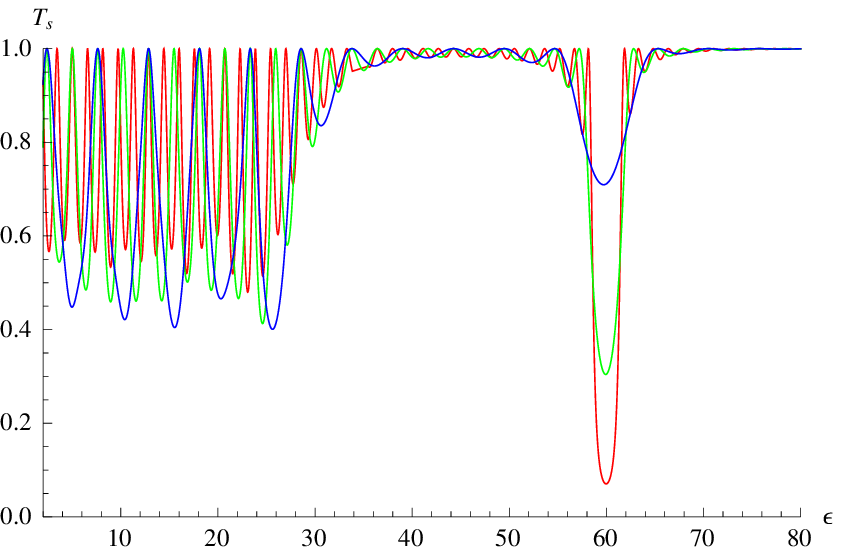}\\
 \caption{\sf{(Color online) (a): GHL shifts $S_t$ in transmission. (b): The transmission probability  
 $T_{s}$ as a function of energy
 $\epsilon$ with  $d_{1}=0.3$ (blue color), $d_{1}=0.6$ (green color),  $d_{1}=1$ (red color), $d_{2}=2.5$, $\mu=0$ and $k_{y}=1$,
  $v_{1}=30$, $v_{2}=60$.}}\lb{fig4}
\end{figure}
In Figure \ref{fig4} we plot the GHL shifts \ref{fig4}(a) and transmission \ref{fig4}(b) as a function of the energy for specific values of the potential parameters $d_{1}=1$,
$d_{2}=2.5$, $k_{y}=1$, $v_{1} =30$ and $v_{2} =60$. It is clear
from Figure \ref{fig4}(a) that the GHL shifts change sign at the Dirac
points ($\epsilon=v_{1}$, $\epsilon = v_{2}$). We deduce
that there is a strong dependence of the GHL shifts on 
$d_{1}$, it increases with $d_{1}$. However, in the energy
domain $\epsilon< v_{1}$ the  GHL shifts 
are positive as long as the energy satisfies the condition
$v_{1}<\epsilon<v_{2}$ and negative for $\epsilon>v_{2}$. We notice that the GHL shifts
display sharp peaks inside the transmission gap around the point
$\epsilon= v_{1}$, while they are absent around the energy point
$\epsilon=v_{2}$. In such situation, one can clearly end up with
an interesting result such that the number of sharp peaks is equal
of that of transmission resonances. We also observe that the
shifts become constant beyond certain energy threshold, which
is compatible with a maximum of transmission.

\begin{figure}[!ht]
\centering
\includegraphics[width=8cm, height=5cm]{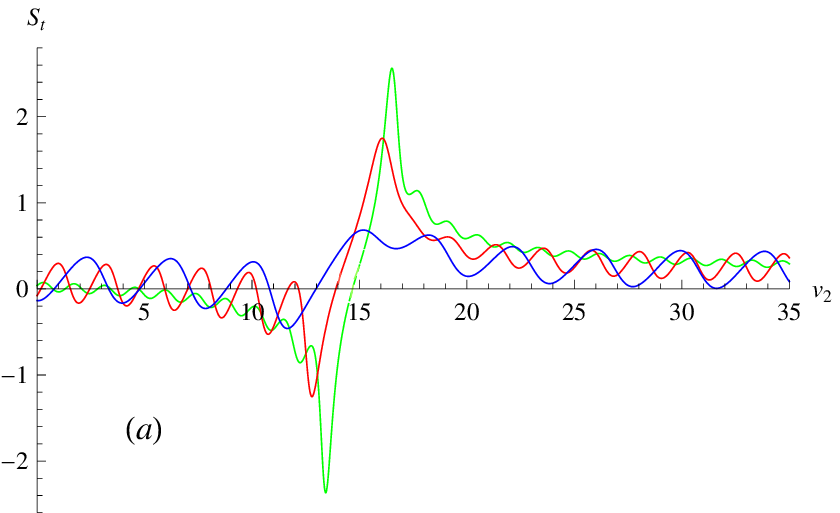}\ \ \ \ \
\includegraphics[width=8cm, height=5cm]{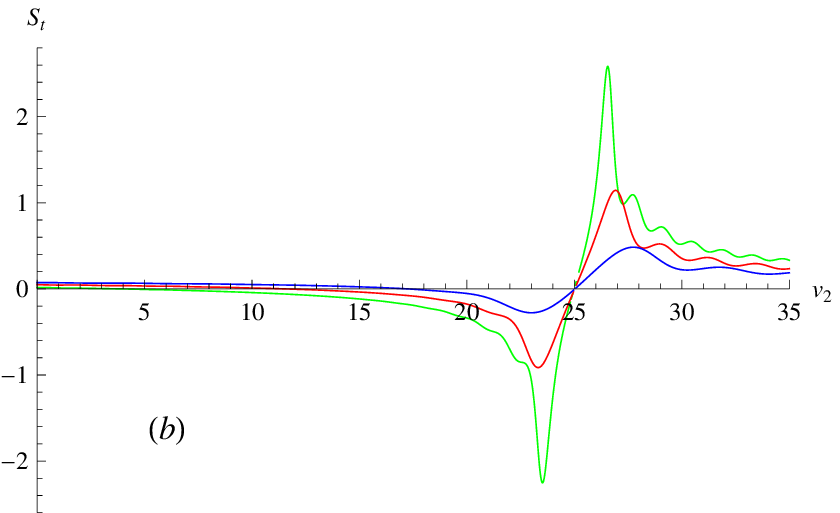}\\
 \caption{\sf{(Color online) GHL shift $S_t$  
 as a function of
 energy potential
 $v_{2}$ with $d_{1}=0.4$ (blue color), $d_{1}=0.7$ (red color), $d_{1}=1.1$ (green color), $d_{2}=1.3$, $\mu=0$,
 $k_{y}=1$,
 $\epsilon=15$ and $v_{1}=30$.}}\lb{figg3}
\end{figure}

 We show in Figure \ref{figg3} the transmission  versus potential strength 
 $v_{2}$. We have chosen the parameters ($\epsilon =15$, $v_{1} = 25$) in
Figure \ref{figg3}(a) and ($\epsilon =25$, $v_{1} = 15$) in Figure \ref{figg3}(b), with inter-barrier
distance $d_2 =1.3$ and distances  $d_1 =\{0.4, 0.7, 1.1\}$. One can notice that, at
the Dirac points $v_{2} = \epsilon$, the GHL shifts change their sign. This change in
sign of the GHL shifts shows clearly that they are strongly
dependent on the barrier heights. We also notice that the GHL
shifts are negative and positive in Figures \ref{figg3}(a) and \ref{figg3}(b). Note
that, the Dirac points represent the zero modes for Dirac operator
\cite{Sharma19} and lead to the emergence of new Dirac points, which  have 
been discussed in different works \cite{Bhattacharjee, Park1}. Such point separates
the two regions of positive and negative refraction. In the cases
of $v_2<\epsilon$ and $v_2>\epsilon$ (respectively $v_{1}< \epsilon$ and $v_1>\epsilon$), the shifts are
respectively in the forward and backward directions, due to the
fact that the signs of group velocity are opposite.

\begin{figure}[!ht]
\centering
\includegraphics[width=8cm, height=5cm]{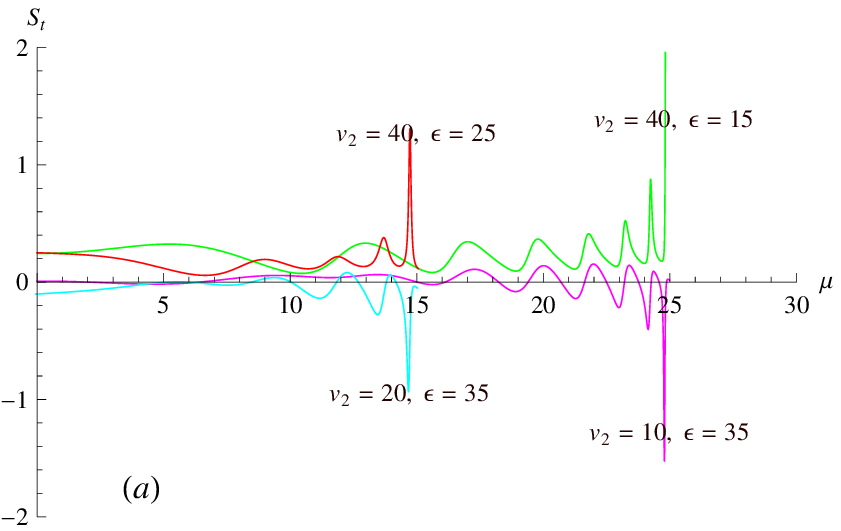}\ \ \ \ \
\includegraphics[width=8cm, height=5cm]{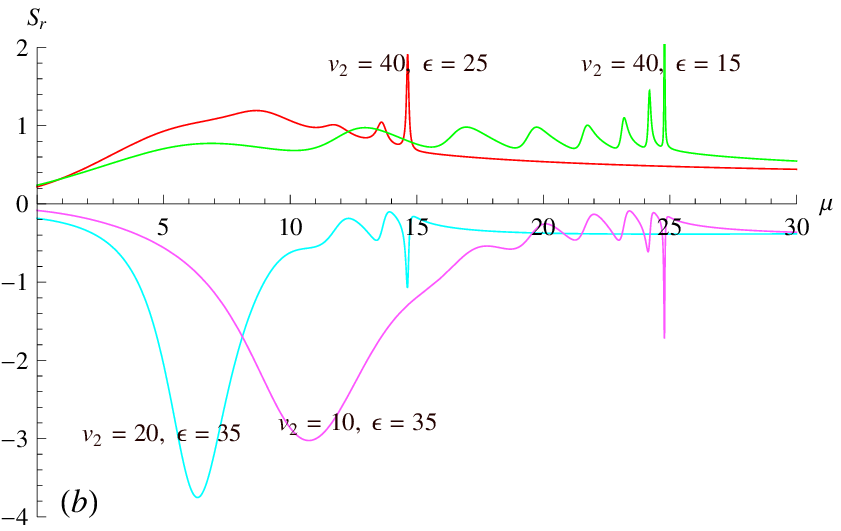}
 \caption{\sf{(Color online) (a)/(b): the GH in transmission and in reflection ($S_{r}/S_{r}$) as a function of
 energy gap $\mu$
  with $d_{1}=0.5$,
 $k_{y}=1$,
 $d_{2}=1.5$ and $v_{1}=50$, $\{v_{2}=40, \epsilon=15\}$, $\{v_{2}=40, \epsilon=25\}$, $\{v_{2}=10, \epsilon=35\}$ and $\{v_{2}=20, \epsilon=35\}$.}}\lb{fig5}
\end{figure}

 Now let us investigate what will happen if we
introduce a gap in the intermediate region $x \leq |d_1| $. Note that, the gap is
introduced as shown in Figure \ref{db.1} and therefore it affects the
system energy according to the solution of the energy spectrum
obtained in region 3. Figure \ref{fig5}(a) and \ref{fig5}(b) show that the
GHL shifts in the propagating case can be enhanced by a gap opening
at the Dirac point. This has been performed by fixing the
parameters $d_{1} =0.5$,  $d_{2} =1.5$, $k_y=1$,  $v_1=50$ and making different
choices for the energy $\epsilon$ and potential $v_{2}$. For
$s_{3}=\mbox{sign}(\epsilon-v_2)=1$ we conclude that one can still have
negative shifts.  Note that \eqref{eq 18i} implies that for
certain energy gap $\mu$, there is no possible transmission. In
fact, under the condition $\mu>|\epsilon-v_{2}|$ every incoming
state is reflected.  We notice that the GHL shifts in transmission $S_{t}$
vanish for values of $\epsilon$ below the critical value
$\mu=|\epsilon-v_{2}|$. Figure \ref{fig5}(b) shows the GHL
 shifts in reflection $S_{r}$ as a function of
 energy gap $\mu$. For the configuration $\{v_2 =40; \epsilon=15, 25\}$,  we can still have
positive shifts while for configuration  $\{\epsilon =35; v_2=10, 20\}$ the GHL shifts are negative.  We notice that the GHL
shifts in reflection $S_{r}$ did not vanish and decreases with increasing $\mu$ for  $s_{3}=\mbox{sign}(\epsilon-v_2)=-1$
as well as increases with increasing $\mu$ for $s_{3}=\mbox{sign}(\epsilon-v_2)=1$.

\begin{figure}[!ht]
\centering
\includegraphics[width=8cm, height=5cm]{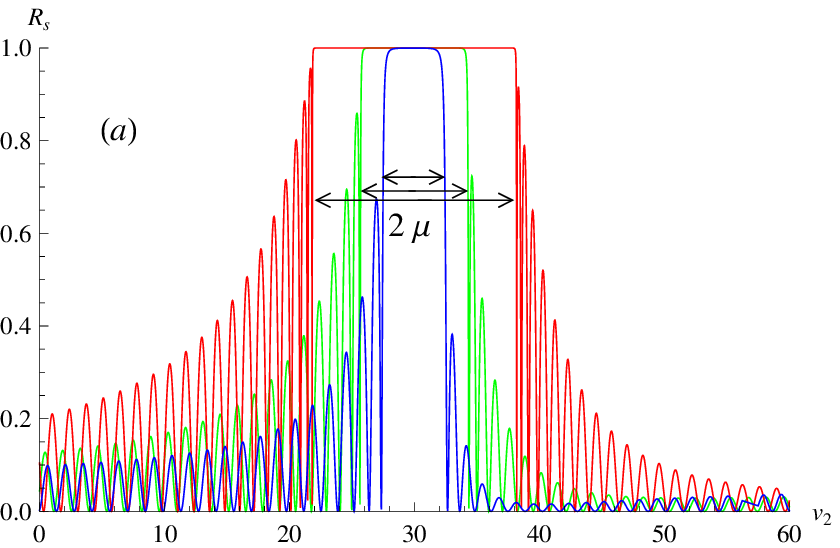}\\
\includegraphics[width=8cm, height=5cm]{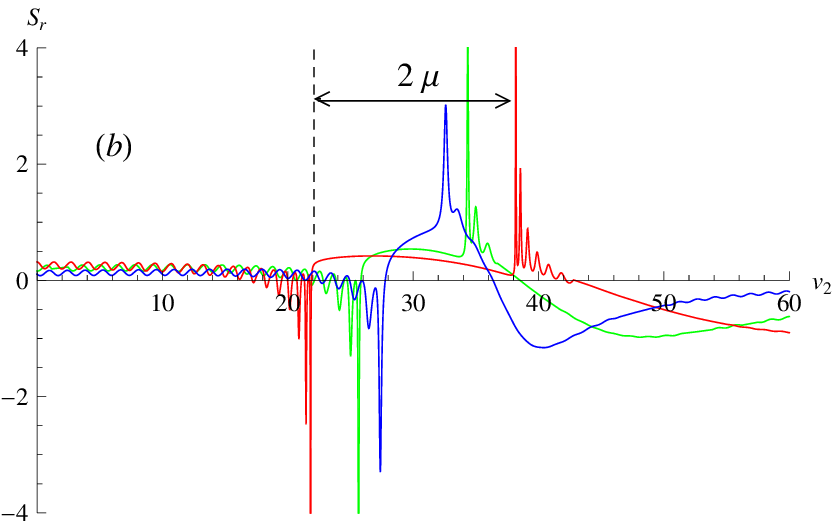}
 \caption{\sf{(Color online) (a): The reflection probability $R_s$ and (b): the GHL shifts  in reflection $S_{r}$ as a function of
 energy gap $\mu$
  with $d_{1}=1.1$,
 $k_{y}=1$,
 $d_{2}=1.3$ and $v_{1}=50$, $\epsilon=35$ and $\mu=2, 4, 8$.}}\lb{fig6}
\end{figure}
The above GHL shifts $S_r$  and reflection probability $R_s$ as function of potential strength $v_2$ for different values of the energy gaps $\mu$ are shown 
in Figure \ref{fig6}(a) and \ref{fig6}(b).
From these Figures, we can see that the region of the
weak GHL shifts become wide with the increase in energy gap $\mu$, the shifts
are affected by the internal structure of the double barrier. In
particular it change the sign at the total reflection
energies and peaks at each bound state associated with the double
linear barrier. Thus the GHL shifts can be
enhanced by the presence of resonant energies in the system when
the incident angle is less than the critical angle associated with
total reflection. It is clearly seen that $S_t$ is oscillating between negative and positive values around the critical point $v_2=\epsilon$. 
At such  point $R_s$ is showing total reflection while it oscillates away from
the critical point.

\begin{figure}[!ht]
\centering
\includegraphics[width=8cm, height=5cm]{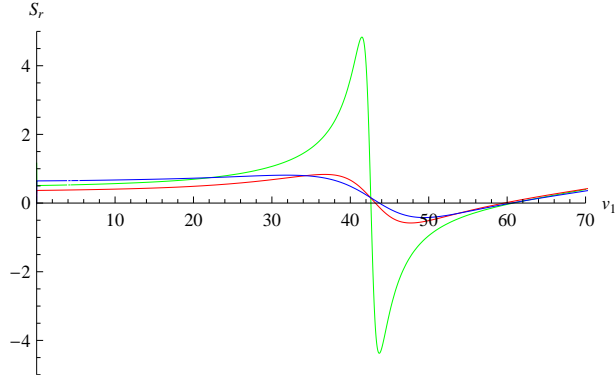}
 \caption{\sf{(Color online) the GHL shifts in reflection ($S_{r}$) as a function of
 energy gap $\mu$
  with $d_{1}=1.1$,
 $k_{y}=1$,
 $d_{2}=1.3$ and $v_{1}=50$, $\epsilon=30$ and $\mu=2, 4, 8$.}}\lb{fig7}
\end{figure}

We show in Figure \ref{fig7} the GHL shifts for the reflection versus potential strength $v_1$ for different energy gaps $\mu$. One observes that the
GHL shifts in reflection can be negative or positive.  Therefore, we can control the positive and negative
GHL shifts by changing the $y$-directional of the energy gap $\mu$. In other words, we can control the moving directions of
the carriers at the interface of the graphene barrier by adjusting $\mu$.

\section{Conclusion}
In this paper we have considered a model to describe over-barrier
electron emission from the edge of monolayer graphene through a
linear electrostatic double barriers.
We have computed the Goos-H\"anchen like (GHL) shifts through a double
barrier potential, the massless Dirac-like equation was used to describe 
the scattered fermions by such a potential configuration.

Our results showed that the GHL shifts are affected by the internal structure 
of the double barrier. In particular the GHL shifts change sign at the transmission zeroes
and peaks at each bound state associated with the double
barrier. Thus our numerical results showed that the GHL shifts can be
enhanced in the presence of resonant energies in the system when
the incident angle is less than the critical angle associated with
total reflection.

Finally we close our work by mentioning some challenges facing the potential connection between
the two fields, quantum optics and graphene. Very recently, pertinent discussions have been made to
emphasis the main difficulties in detecting the GHL shifts and preparing the electron beam
in solid-state physics \cite{xxChen}. These discussions open for us important research avenues that will help us
understand and overcome the above mentioned difficulties. On the other hand, we learned from \cite{xxChen}
that the spin-orbit coupling in optics is an interesting and fascinating topic because the spin-orbit
interaction in graphene opens up a spin-orbit gap, though very small, at the Dirac points. All these
matters will be highly important when we consider tunable GH shift leading to potential applications
in future graphene based electronic devices. These matters will be investigated in the near future to
enable us to get a deeper understanding of graphene transport properties.

\section*{Acknowledgments}

The generous support provided by the Saudi Center for Theoretical
Physics (SCTP) is highly appreciated by all authors. AH and HB
acknowledge the support of King Fahd University of Petroleum and
minerals under research group project entitled 
"Transport Properties of Systems Composed of Coupled and Decoupled Graphene Flakes".



\begin{thebibliography}{99}

\bibitem{Novoselov} K. S. Novoselov, A. K. Geim, S. V. Morozov, D. Jiang, Y. Zhang, S. V.
Dubonos, I. V. Grigorieva and A. A. Firsov, {Science} {306}, 666
(2004).

\bibitem{FGoos}  F. Goos and H. H\"anchen, Ann. Phys. 436, 333 (1947).

\bibitem{FFGoos}  F. Goos and H. H\"anchen, Ann. Phys. 6, 251 (1949).

\bibitem{Artmann}  K. Artmann, Ann. Phusik 2, 87 (1949).


\bibitem{Chen15} X. Chen, J.-W. Tao and Y. Ban, Eur. Phys. J. B 79, 203
(2011).

\bibitem{Song16} Y. Song, H-C. Wu and Y. Guo, Appl. Phys. Lett. 100, 253116
(2012).

\bibitem{Chen18} X. Chen, P-L. Zhao, X-J. Lu and L-G. Wang, Eur. Phys. J. B
86, 223 (2013).

\bibitem{Sharma19} M. Sharma and S. J. Ghosh, J. Phys.: Condens. Matter 23, 055501
(2011).

\bibitem{Huang13} J.-H. Huang, Z.-L. Duan, H.-Y. Ling and W.-P. Zhang, Phys. Rev. A
77, 063608 (2008).

\bibitem{Beenakker}  C. W. J. Beenakker, R. A. Sepkhanov, A. R. Akhmerov and J. Tworzydlo, Phys. Rev. Lett. 102, 146804 (2009).


\bibitem{Zhao11}  L. Zhao and S. F. Yelin, Phys. Rev. B 81, 115441 (2010).


\bibitem{MMekkaoui} M.
Mekkaoui, A. Jellal and H. Bahlouli,  arXiv:1511.06880 (2015).


 \bibitem{Bohm} D. Bohm,
Quantum Theory, Prentice-Hall (New York, 1951), pp. 257-261.


\bibitem{AAJellal}  A. Jellal, I. Redouani, Y. Zahidi and H. Bahlouli, Phys. E 58, 30 (2014).

\bibitem{AAAJellal}  A. Jellal, Y. Wang, Y. Zahidi and M. Mekkaoui, Phys. E 68, 53 (2015).

\bibitem{Matulis} A. Matulis, F. M. Peeters and P. Vasilopoulos, Phys. Rev. Lett.
72, 1518 (1994).

\bibitem{Ramezani} M. Ramezani Masir, P. Vasilopoulos and F.M. Peeters, Phys.
Rev. B 82, 115417 (2010).

\bibitem{Tworzydlo} J. Tworzydlo, B. Trauzettel, M. Titov, A. Rycerz and C. W. J. Beenakker, Phys. Rev. Lett. 96,
246802 (2006).





\bibitem{Abramowitz}  M. Abramowitz and I. Stegum, Handbook of Integrabls, Se ries and
Products, (Dover, New York, 1956).

\bibitem{Gonzalez} L. Gonzalez-Diaz and V. M. Villalba, Phys. Lett. A 352, 202
(2006).

\bibitem{Bhattacharjee} S. Bhattacharjee, M. Maiti and K. Sengupta, Phys. Rev. B 76, 184514 (2007).

\bibitem{Park1} C. H. Park, L. Yang, Y. W. Son, M. L. Cohen and S. G.Louie, Phys. Rev. Lett. 101, 126804 (2008).

\bibitem{xxChen} X. Chen, X.-J. Lu, Y. Ban and C.-F. Li, J. Opt. 15,
033001 (2013).








\bibitem{HBahlouli} H. Bahlouli, E.B. Choubabi, A. EL Mouhafid and A. Jellal, Solid
State Communications 151 (2011) 1309.


\bibitem{Berry} M. V. Berry and R. J. Modragon, Proc. R. Soc. London Ser. A 412, 53 (1987).


\end{thebibliography}
\end{document}